\documentclass[%
 reprint,
superscriptaddress,
showpacs,preprintnumbers,
 amsmath,amssymb,
 aps,nolinenumbers
]{revtex4-1}
\usepackage{ulem}
\usepackage{ragged2e}
\usepackage{caption}
\usepackage{graphicx}
\usepackage{dcolumn}
\usepackage{bm}
\usepackage{xcolor}

\begin{document}
\captionsetup{justification=centerlast , labelsep = space,labelfont=bf,format=plain,singlelinecheck=off}

\title{Fragile pressure-induced magnetism in FeSe superconductors\\ with a thickness reduction}

\author{Jianyu~Xie$^\ddagger$}
\author{Xinyou~Liu$^\ddagger$}
\author{Wei~Zhang}
\author{Sum~Ming~Wong}
\affiliation{Department of Physics, The Chinese University of Hong Kong, Hong Kong SAR, China}
\author{Xuefeng~Zhou}
\author{Yusheng~Zhao}
\author{Shanmin~Wang}
\affiliation{Department of Physics, Southern University of Science and Technology, Shenzhen, Guangdong, China}
\author{Kwing~To~Lai$^{*}$}
\author{Swee K. Goh$^{*},$}
\affiliation{Department of Physics, The Chinese University of Hong Kong, Hong Kong SAR, China}

\date{September 1, 2021}

\begin{abstract}

The emergence of high transition temperature ($T_c$) superconductivity in bulk FeSe under pressure is associated with the tuning of nematicity and magnetism. However, sorting out the relative contributions from magnetic and nematic fluctuations to the enhancement of $T_c$ remains challenging. Here, we design and conduct a series of high-pressure experiments on FeSe thin flakes. We find that, as the thickness decreases, the nematic phase boundary on temperature-pressure phase diagrams remains robust while the magnetic order is significantly weakened. A local maximum of $T_c$ is observed outside the nematic phase region, not far from the extrapolated nematic endpoint in all samples. However, the maximum $T_c$ value is reduced associated with the weakening of magnetism. No high-$T_c$ phase is observed in the thinnest sample. Our results strongly suggest that nematic fluctuations alone can only have a limited effect while magnetic fluctuations are pivotal on the enhancement of $T_c$ in FeSe. 

\end{abstract}


\maketitle

Unconventional superconductors including high-$T_c$ cuprates, heavy-fermion systems, and iron-based superconductors (IBS) share common features in the intimate relations with magnetic, structural, and electronic orders~\cite{ Gegenwart2008, si2010heavy, Sato2017, Mathur1998, Imai2009, Lai2014, Kamihara2008, Keimer2015, Kasahara2010, Bohmer2017, Shibauchi2014,  Kawashima2021}. Understanding the impact of these orders on the stabilization or the optimization of superconductivity is a central topic. Using tuning parameters such as chemical substitution and high pressure, these ordered phases can be controlled, providing a means to disentangle their role on the emergence of superconductivity. In some IBS, such as Ba(Fe$_{1-x}$Co$_x$)$_2$As$_2$, the emergence of the antiferromagnetic (AFM) ordered phase or spin density wave (SDW) phase is accompanied by a nematic phase transition~\cite{Chu2010,Fernandes2014}, in which the rotational symmetry is broken but the translational symmetry is preserved. When the nematic and SDW phase transitions are sufficiently weakened, the superconductivity appears and develops into a dome shape. Recent studies suggest that, near the quantum critical points of nematicity and SDW order, both enhanced nematic fluctuations and magnetic fluctuations can mediate Cooper pairing~\cite{Kontani2010, Lederer2015, Lederer2017, Shibauchi2014}. However, due to the proximity of the nematic phase and the magnetically ordered phase, the interplay among superconductivity, nematicity, and magnetism remains elusive.\par

FeSe, as another prominent member of IBS, has drawn intensive attention~\cite{Imai2009,Bohmer2017,Margadonna2008,Mcqueen2009,Baek2015,Bohmer2013,Bohmer2015,Hsu2008,Fernandes2014,Gati2019,Kang2020,Ying2018,Shi2018,Kothapalli2016,Kaluarachchi2016,Sun2016,Sun2017,Li2017,He2013,Tan2013,Ge2015,Farrar2020,Lei2016,Lei2017,Miyata2015,Shiogai2016,Bohmer2016,Zhang2016, Huang2016,Margadonna2008,Bendele2012,Hosoi2016,Matsuura2017,Yip2017,Sato2018,Reiss2020,licciardello2019,Huang2020,Coldea2019,Chibani2021,Rana2020,Mukasa2021, Ge2019, Zhang2021}. Bulk FeSe undergoes a nematic phase transition at $T_s\sim90$~K, and further becomes superconducting at $T_c=9$~K, without the presence of magnetism~\cite{Margadonna2008, Mcqueen2009, Imai2009, Baek2015,Bohmer2013, Bohmer2015, Hsu2008}. Under pressure, the SDW phase can be induced at around 10~kbar~\cite{Kothapalli2016, Kaluarachchi2016, Bendele2012}, which develops into a dome shape in the temperature-pressure ($T$-$p$) phase diagram and spans more than 40~kbar~\cite{Sun2016,Sun2017}. Concomitantly, the superconductivity is enhanced in a three-step process and reaches $\sim40$~K at around 60~kbar. The novel phases in FeSe can also be tuned by combining chemical substitution and pressure~\cite{Hosoi2016, Matsuura2017, Yip2017, Sato2018, Reiss2020, licciardello2019, Huang2020, Li2017, Coldea2019}. In FeSe$_{1-x}$S$_x$, the nematic phase is suppressed as sulfur content increases, and a nematic quantum critical point is observed without the presence of magnetism~\cite{Hosoi2016,licciardello2019,Chibani2021}. Meanwhile, the SDW phase in FeSe$_{1-x}$S$_x$ only appears when pressure is applied and its onset pressure shifts to a higher value as sulfur content increases~\cite{Matsuura2017, Rana2020, Reiss2020}. Intriguingly, high-$T_c$ phases appear at both ends of the SDW dome, suggesting that the enhanced magnetic fluctuations may help to increase $T_c$~\cite{Matsuura2017}. A recent result on FeSe$_{1-x}$Te$_x$ shows that, as Te content increases, the nematic phase can also be suppressed, while the pressure-induced SDW phase disappears gradually against $x$~\cite{Mukasa2021}. At $x>0.10$, no magnetic order is observed under pressure~\cite{Mukasa2021}. However, a superconducting dome still forms near the nematic phase, indicating that at least for the FeSe$_{1-x}$Te$_x$ series, the nematic fluctuations alone may also help to promote $T_c$~\cite{ Mukasa2021}. \par

In addition to pressure and chemical substitution, the superconductivity in FeSe can also be tuned by sample thickness~\cite{He2013, Tan2013, Ge2015, Farrar2020, Lei2016, Lei2017, Ying2018, Miyata2015, Shiogai2016}. Monolayer FeSe grown on SrTiO$_3$ substrate exhibits a remarkably high $T_c$ over 100~K~\cite{Ge2015}, much higher than in the bulk sample under pressure. On the other hand, thin FeSe films cleaved from single crystals exhibit suppression of $T_c$ and $T_s$ as the thickness is reduced~\cite{Farrar2020, Lei2016, Lei2017, Ying2018}. Since the magnetically ordered phase in the bulk FeSe only appears with pressure applied, whether thickness reduction alters the magnetism in thin FeSe remains unknown yet an important topic. 
According to Mermin–Wagner theorem, the long-range order with broken symmetry is prohibited in two-dimensional materials~\cite{Mermin1966}. Therefore, with a decreasing thickness, the magnetic order is expected to be weakened, paving a way to critically examine the bordering high-$T_c$ superconductivity. Thus, the ability to simultaneously apply pressure and to alter the sample thickness provides a powerful tool to interrogate the relationship between magnetism and superconductivity in FeSe.

\begin{figure}[!t]\centering
      \resizebox{8.5cm}{!}{
              \includegraphics{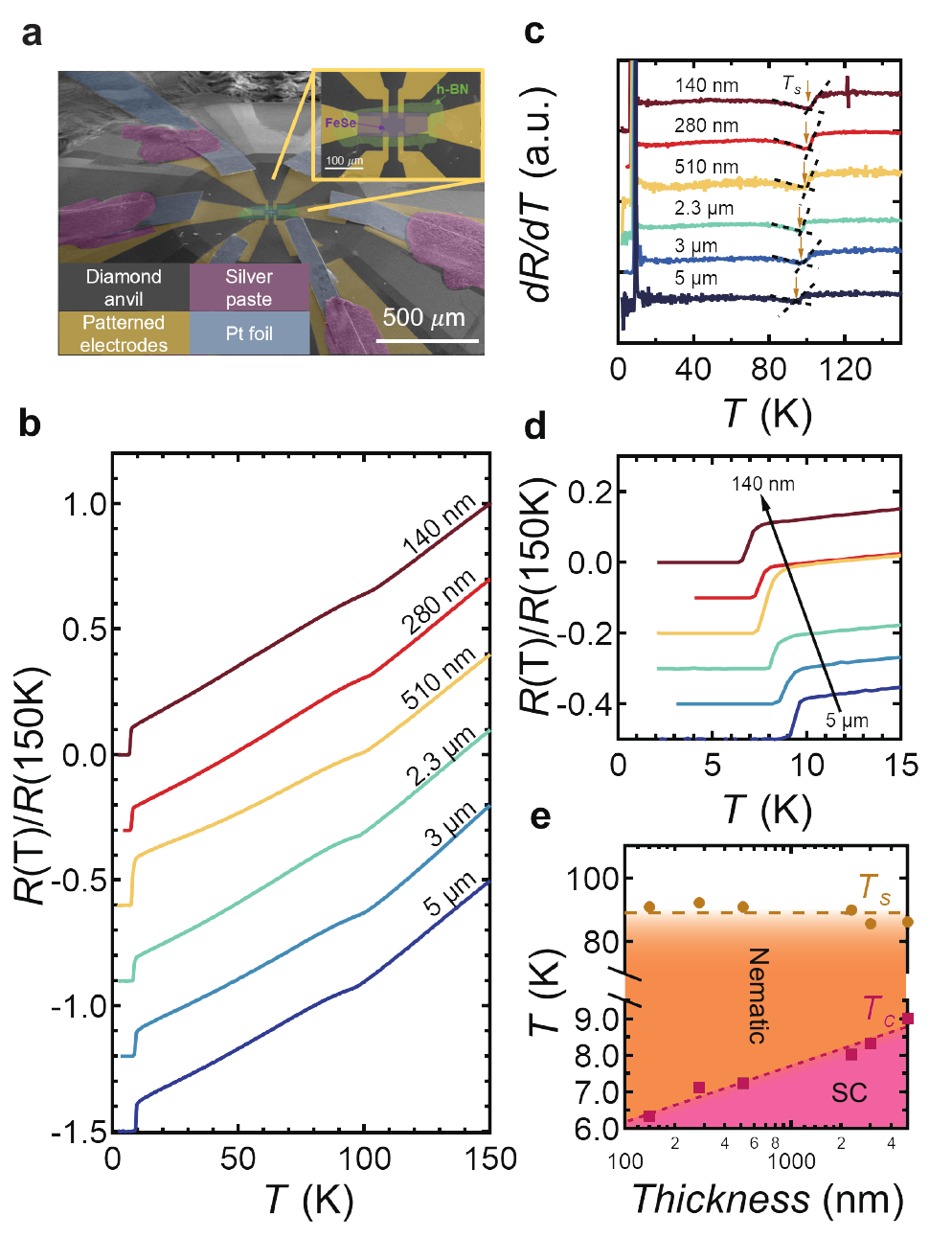}}                				
              \caption{\label{fig1} 
              \textbf{(a)} The scanning electron microscope (SEM) image of the device integrated diamond anvil. The inset displays the top view of a 280~nm thick FeSe sample covered with thin h-BN. False colors are used for illustration. \textbf{(b)} The temperature dependence of normalized resistance, $R/R(150~{\rm K})$, for samples with different thicknesses, collected on warming. \textbf{(c)} The temperature dependence of $dR/dT$ for the curves in \textbf{b}. The arrows indicate the temperature of the nematic phase transition. \textbf{(d)} The temperature dependence of $R/R(150~{\rm K})$ at $T<15$~K. The curves in \textbf{b-d} are vertically shifted for clarity. \textbf{(e)} The temperature-thickness phase diagram of measured thin FeSe. $T_s$ (brown) is determined from both warming and cooling data in $dR/dT$ (see Supporting Information~S2). $T_c$ is identified as the temperature at which the resistance reaches zero. The regions for nematic, and superconducting phases are coloured for \leftline{clarity.}
              }
\end{figure}

Here, we report our high-pressure investigation of superconducting, nematic, and magnetic phase transitions in FeSe with thicknesses ranging from 140~nm to 5~$\mu$m using the concept of a ``device-integrated diamond anvil cell technique" we developed (see Methods). Thin flakes of FeSe are first isolated from high-quality single crystals and then they are transferred onto the diamond anvils pre-patterned with electrodes. The diamond anvil integrated with thin FeSe devices (see Fig.~\ref{fig1}(a)) then forms a part of the diamond anvil cell for resistance measurements under hydrostatic pressure. We find that, as the thickness is reduced, the nematic phase shows similar pressure evolutions, while the SDW phase shrinks significantly. Importantly, the high-$T_c$ superconductivity can only be observed in 
thicker samples, in which the SDW phase persists to higher pressure and temperature. In other words, the robustness of the SDW phase is crucial for stabilizing the high-$T_c$ superconductivity in FeSe. Using the constructed $T$-$p$ phase diagrams, we extract the interplay among nematicity, magnetism, and superconductivity.


Figure~\ref{fig1}(b) displays the electrical resistance ($R$) as a function of temperature $(T)$ normalized to $R(150~{\rm K})$ for samples with the thickness ranging from 140~nm to 5~$\mu$m, collected on warming at ambient pressure. At around 100~K, pronounced slope changes due to the nematic phase transition are observed in all samples. Here, we define the nematic phase transition temperature $T_s$ as the temperature at which the $dR/dT$ is a local minimum (see Fig.~\ref{fig1}(c)). To alleviate the effect of the thermal lag due to the usage of a pressure cell, both warm-up or cool-down data are analyzed for extracting $T_s$ accurately (see Supporting Information~S2).\par

Figure~\ref{fig1}(d) displays the normalized resistance $R(T)/R(150~{\rm K})$ at $T<15$~K. Like the bulk sample~\cite{Margadonna2008, Mcqueen2009, Imai2009, Baek2015, Bohmer2015, Hsu2008}, thin FeSe samples undergo a superconducting transition at low temperatures. However, a systematic suppression of $T_c$, defined as the temperature at which the resistance reaches zero, is observed as the thickness is reduced. Together with the $T_s$ data, the temperature-thickness phase diagram at ambient pressure is shown in Fig.~\ref{fig1}(e). In the thickness range explored, our $T_s$ of around 90~K agrees quantitatively with previous results on the bulk FeSe~\cite{ Margadonna2008, Mcqueen2009, Imai2009, Baek2015, Bohmer2015, Hsu2008}. 
Considering that the superconducting transitions observed are sharp (Fig.~\ref{fig1}(d)) and the suppression of $T_c$ against thickness is systematic, we conclude that the observed phenomena are the intrinsic properties from thickness reduction.

\begin{figure*}[!t]\centering
      \resizebox{18cm}{!}{
              \includegraphics{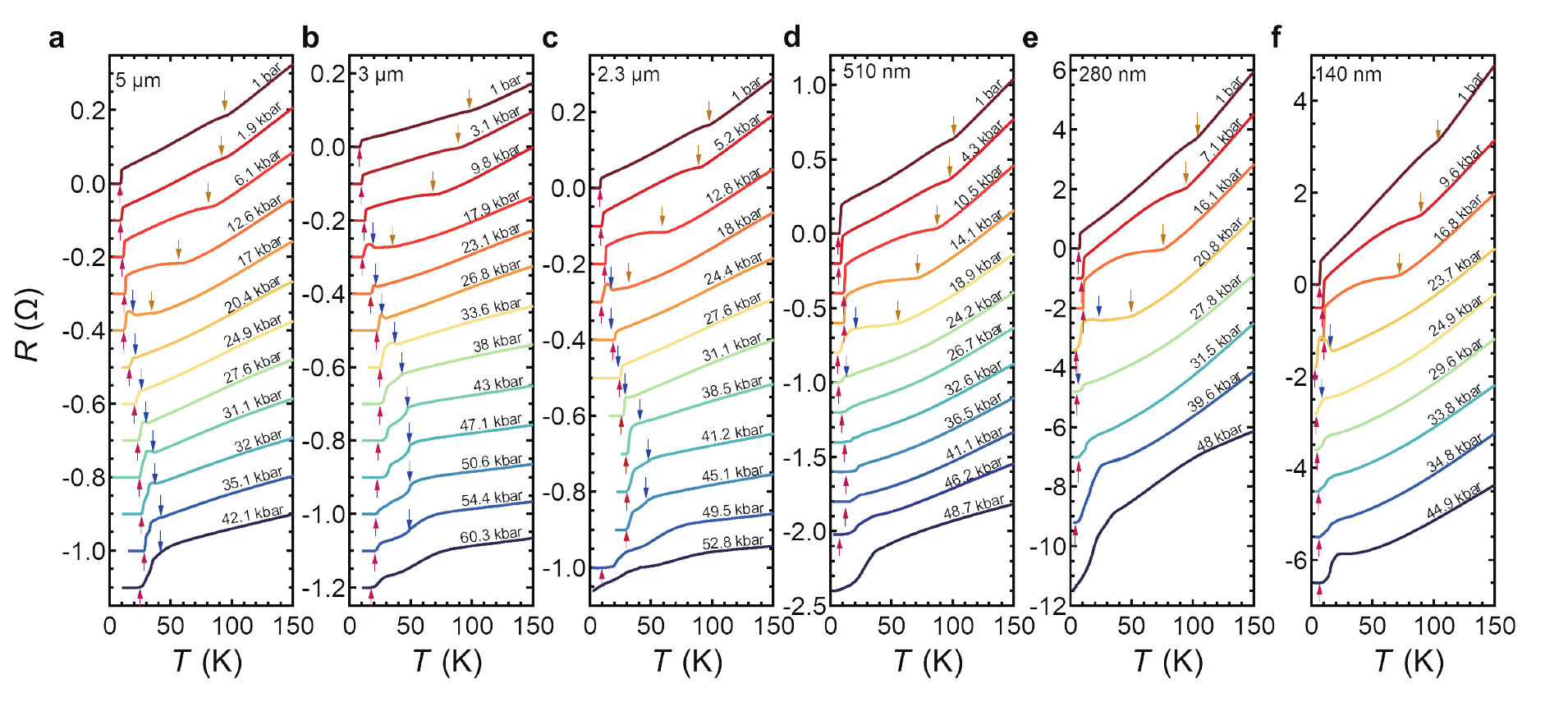}}                				
              \caption{\label{fig2} 
              The temperature dependence of resistance, $R(T)$, collected on warming for thin FeSe under high pressure with thickness of \textbf{(a)} 5~$\mu$m, \textbf{(b)} 3~$\mu$m, \textbf{(c)} 2.3~$\mu$m, \textbf{(d)} 510~nm, \textbf{(e)} 280~nm and \textbf{(f)} 140~nm. The anomalies in $R(T)$ corresponding to $T_c$, $T_s$, and $T_m$ are indicated by red, brown and blue arrows, respectively. Except for the ambient \leftline{pressure data, the curves are vertically shifted for clarity.} } 
              
\end{figure*}
\begin{figure}[htpb]\centering
      \resizebox{8cm}{!}{
              \includegraphics{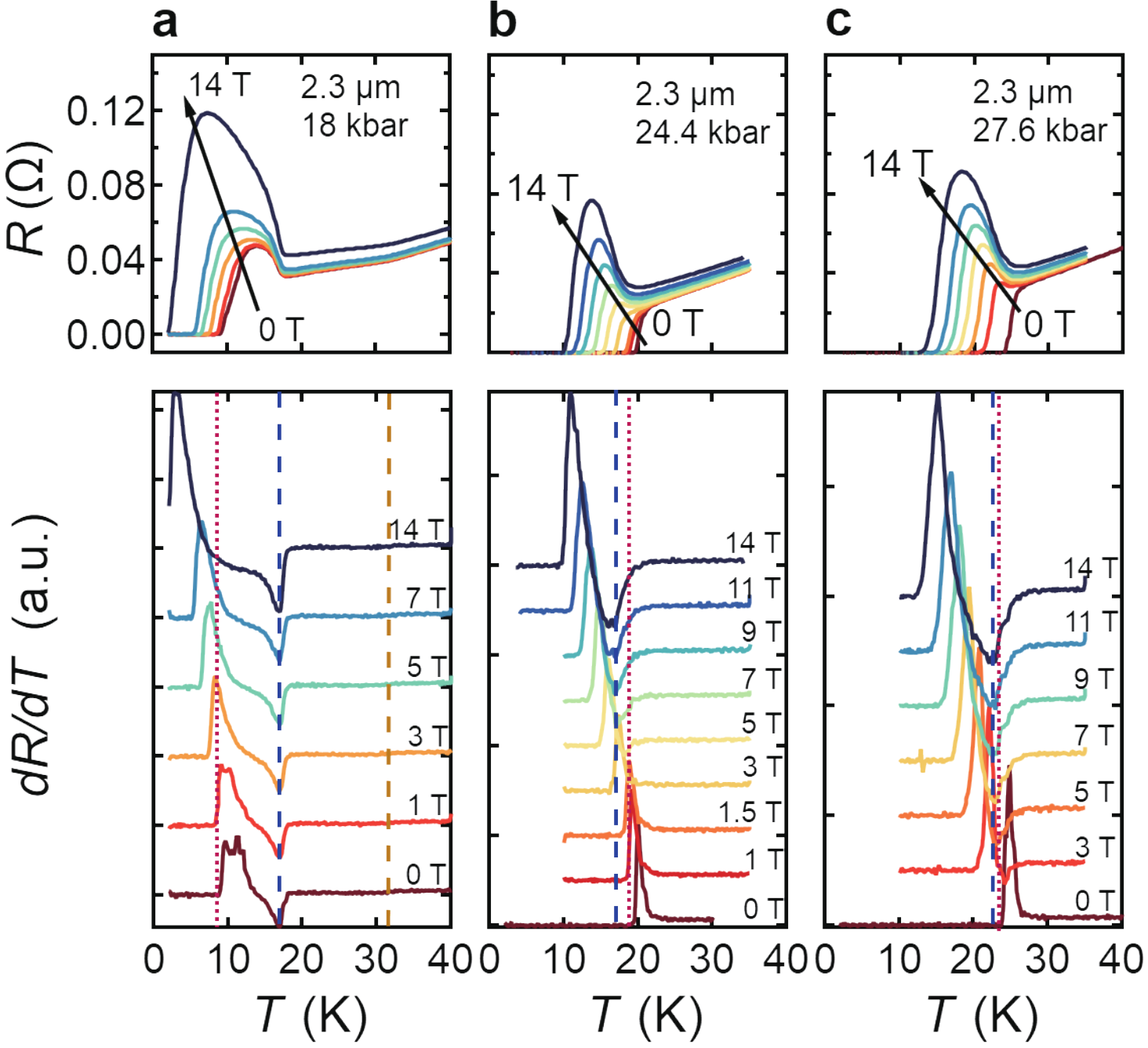}}                				
              \caption{\label{fig3} 
              The temperature dependence of resistance (upper) and its derivative (lower) in fields up to 14~T with $I//ab\perp B$ for the 2.3~$\mu$m thick sample at \textbf{(a)} 18~kbar, \textbf{(b)} 24.4~kbar and \textbf{(c)} 27.6~kbar. Zero field $T_c$ are indicated by the vertical red dotted lines. $T_s$ and $T_m$ are indicated by the vertical dashed brown and dashed blue \leftline{lines, respectively. }
              }
\end{figure}
\begin{figure*}[htpb]\centering
      \resizebox{14cm}{!}{
               \includegraphics{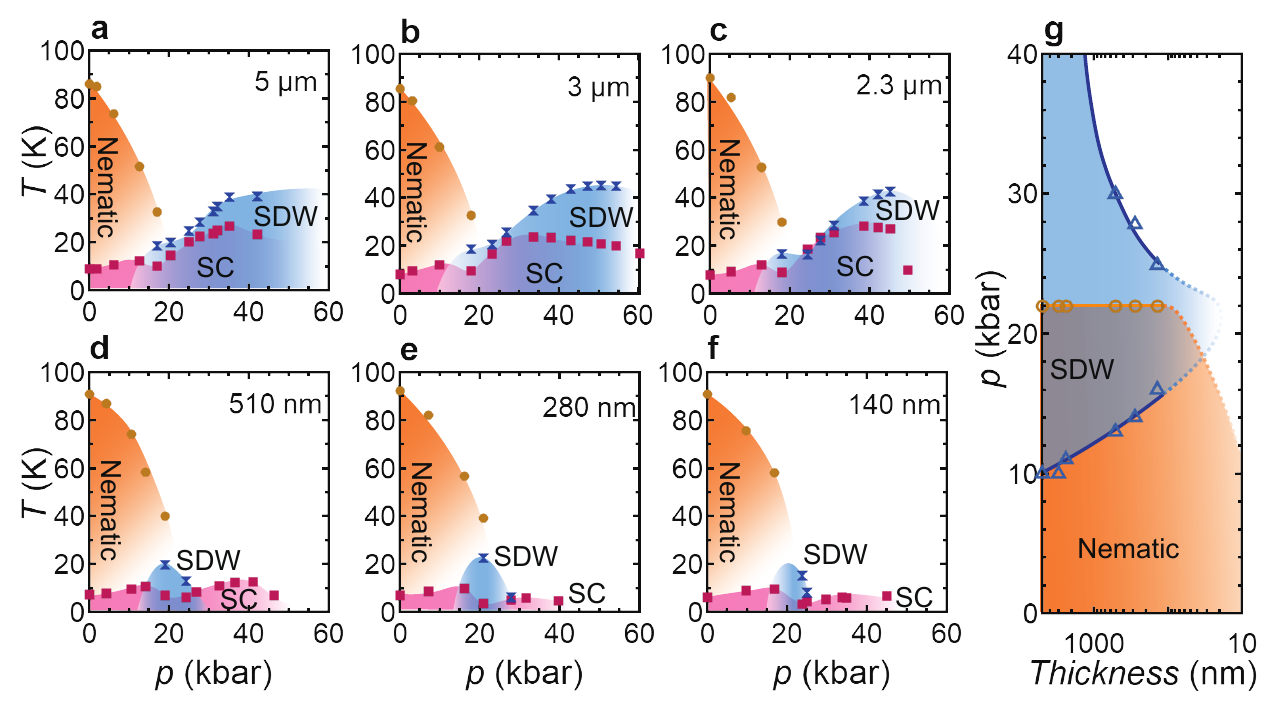}}                				
              \caption{\label{fig4} 
              The $T$-$p$ phase diagram in FeSe with thickness of \textbf{(a)} 5~$\mu$m, \textbf{(b)} 3~$\mu$m, \textbf{(c)} 2.3~$\mu$m, \textbf{(d)} 510~nm, \textbf{(e)} 280~nm and \textbf{(f)} 140~nm. The phase transition temperatures $T_c$ (purple), $T_s$ (brown) and $T_m$ (blue) are determined by the anomalies in $R(T)$ and $dR/dT$ (see Fig.~\ref{fig2}, Fig.~\ref{fig3}, Supporting Information~S2 and Supporting Information~S3). The nematic, SDW and superconducting phases are coloured for clarity. \textbf{(g)} The schematic pressure-thickness phase diagram at $T=0$~K proposed based on the results in \textbf{a-e} and Refs.~\cite{Sun2016, Bendele2012, Kothapalli2016}. The symbols represent the extrapolated phase boundary at the $T=0$~K limit. The dashed orange curve is drawn based \leftline{on the expectation that $T_s$ would vanish at a lower pressure when the ambient pressure $T_s$ is lower.}
              }
\end{figure*}

Figure~\ref{fig2} displays the evolution of $R(T)$ under pressure for samples with thicknesses of 5~$\mu$m, 3~$\mu$m, 2.3~$\mu$m, 510~nm, 280~nm, and 140~nm. As pressure increases, the anomaly in $R(T)$ corresponding to the nematic phase transition shifts to a lower temperature, as indicated by the brown arrows. In Supporting Information~S3, we show the evolution of $dR/dT$ under pressure for illustration. At $p>20$~kbar, no feature corresponding to the $T_s$ is observed in $R(T)$ or $dR/dT$, suggesting the complete suppression of the nematic phase. Our observations also agree well with previous results in bulk FeSe under high pressure~\cite{Bendele2012, Kothapalli2016, Sun2016}. Given that the thin FeSe samples in this study possess similar $T_s$ at ambient pressure, the consistency in the pressure evolution of $T_s$ in samples with different thicknesses is not surprising. This observation suggests that, in the thickness range investigated, the nematic quantum criticality and nematic fluctuations may not be substantially affected by the thickness reduction.

Next we turn to the investigation of the pressure-induced SDW phase. Since the 5~$\mu$m and 3~$\mu$m thick samples exhibit similar behaviors under pressure, we choose the data in 3~$\mu$m thick sample for a detailed discussion. At 17.9~kbar, an additional anomaly manifested as a pronounced upturn is observed in $R(T)$ at 19.1~K (blue arrow). As pressure increases, this anomaly gradually shifts to higher temperatures. At $p\geqslant38$~kbar, the anomaly becomes a downward kink. 
The evolution of this $R(T)$ anomaly under pressure agrees well with the assignments of the SDW phase transition in $R(T)$ reported previously in bulk FeSe, FeSe$_{1-x}$S$_x$ series and FeSe$_{1-x}$Te$_x$ series under pressure~\cite{Sun2016, Matsuura2017, Mukasa2021}. Following the same procedure~\cite{Sun2016}, we obtained the SDW phase transition temperature $T_m$ from either the peak or dip features in $dR/dT$~(see Supporting Information~S3). 

In 2.3~$\mu$m thick sample, a pronounced upturn is also observed in $R(T)$ at 18~kbar. At 24.4~kbar, and 27.6~kbar, sharp superconducting transitions are observed, while no anomaly corresponding to the SDW phase transition is observed. Inspired by the results on bulk FeSe at 60~kbar, in which the anomaly associated with the SDW phase transition emerges when the superconductivity is suppressed by the high magnetic fields~\cite{Sun2016}, we conducted magneto-electrical resistance measurements. Figure~\ref{fig3} displays the obtained results with $I\| ab\perp B$. At 18~kbar, the upturn becomes even more pronounced at high fields. The enhanced magnetoresistance in the SDW phase suggests that the Fermi surfaces or/and carrier scattering rates are affected by the pressure-induced antiferromagnetism. At 24.4~kbar and 27.6~kbar, pronounced upturns appears in the $R(T)$ at high fields. The corresponding dip features in $dR/dT$ become visible when the field is larger than 7~T, and remain robust up to 14~T, satisfying the signature of the magnetic phase transition~\cite{Sun2016}. As pressure continues to increase, the anomaly associated with $T_m$ shows up again in the zero-field $R(T)$ curves, and reaches over 40~K at $p\sim$40~kbar. At higher pressures, $T_m$ can no longer be unambiguously assigned. \par

In 510~nm, 280~nm and 140~nm thick samples, the pressure evolution of the SDW phase shows a drastically different behaviour compared with the few-micrometer thick samples. At $\sim$20~kbar, additional upturn can be observed in $R(T)$ at zero field (see Fig.~\ref{fig2}). Utilizing the same strategy of measuring $R(T)$ in high fields, we also confirmed the SDW phase transition in the 280~nm thick sample at 27.8~kbar, and in the 140~nm thick sample at 24.9~kbar (see Supporting Information~S3). At $p>30$~kbar, no SDW phase transition is observed for all three samples with thicknesses of a few hundred nanometers. Hence, the SDW region shrinks substantially in the $T$-$p$ phase diagrams, indicating a rapid weakening of magnetism in these thin FeSe samples. \par

Finally, we turn to the pressure evolution of $T_c$ in thin FeSe. In all samples measured, $T_c$ first experiences an enhancement at low pressures ($p<\sim15$~kbar), and then it is suppressed when approaching the critical pressure of the nematic phase transition ($p\approx20$~kbar). The situation is more complex away from the nematic phase. For the group of thicker samples, namely 5~$\mu$m, 3~$\mu$m, and 2.3~$\mu$m thick samples, 
$T_c$ is enhanced initially with the accompanying enhancement of $T_m$, and $T_c$ reaches 27.0~K at 35.1~kbar for 5~$\mu$m thick sample, 23.7~K at 33.6~kbar for 3~$\mu$m thick sample, and 28.2~K at 38.5~kbar for 2.3~$\mu$m thick sample. At higher pressure, $T_m$ continues to increase and saturates, while $T_c$ begins to drop. In the thinner group, namely the 510~nm, 280~nm, and 140~nm thick samples, $T_c$ only experiences a slight increase after leaving the nematic phase. In the 510~nm thick sample, $T_c$ reaches 12.6~K at 41.1~kbar, similar to the peak value of $T_c$ when the nematic phase is present. At 48.7~kbar, no zero resistance is observed down to 2~K. In 280~nm and 140~nm thick samples, $T_c$ values measured are less than 7~K, below the maximum $T_c$ values achieved in the nematic phase. Considering the SDW phase is fragile in the thinner group, the suppression of $T_c$ can be naturally associated with the weakening of the magnetism.

In contrast to the thickness dependence of physical properties in many 2D materials~\cite{Li2019,sajadi2018gate,fatemi2018electrically}, a reduction of thickness to hundreds of nanometers can already alter the physical properties of FeSe under pressure.
Based on our $T$-$p$ phase diagrams and previous results~\cite{Sun2016, Bendele2012, Kothapalli2016, Farrar2020}, we propose a schematic pressure-thickness phase diagram at $T=0$~K showing the nematic and SDW phase boundaries, as shown in Fig.~\ref{fig4}(g).
From the schematic phase diagram, it can be seen that the nematic phase boundary is nearly unaffected by the thickness reduction over a wide range of thicknesses, as shown by the nearly horizontal phase boundary above $\sim100$~nm.
On the other hand, when the thickness is reduced, the SDW phase becomes more fragile, and the pressure range over which the SDW phase exists shrinks significantly. As presented above, the weakening of the SDW order has an observable effect on the reduction of $T_c$.

\begin{figure}[!t]\centering
      \resizebox{8.5cm}{!}{
              \includegraphics{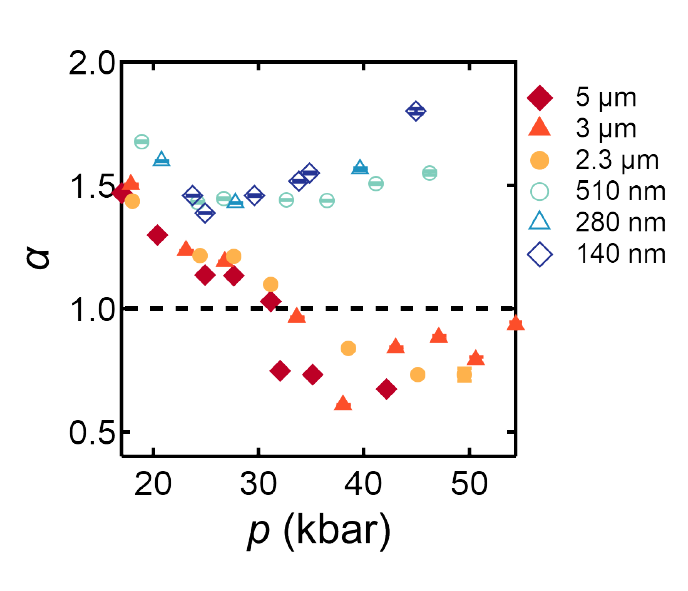}}
              \caption{\label{fig5} 
              The temperature exponent $\alpha$ for the resistance as a function of pressure for $p$\hspace{0.21111em}$>$\hspace{0.21111em}$16$~kbar and $T$ between 100~K and max\{$T_m$, $T_s$\}. For detailed data analysis \leftline{procedure, see Supporting Information~S4.}}  
              
\end{figure}

Why is a robust SDW phase crucial for realizing high-$T_c$ superconductivity in FeSe? In superconductors driven by magnetic quantum criticality, the superconductivity is optimized near the magnetic quantum critical point, at which magnetic fluctuations are critically enhanced ~\cite{Shibauchi2014, Kasahara2012, Mathur1998, Gegenwart2008}. Such a scenario highlights the importance of magnetic fluctuations as an effective pairing glue. Thus, it is natural to speculate that magnetic fluctuations are debilitated in the thinner group of samples. 

To explore this hypothesis further, we examine the zero-field $R(T)$ between max\{$T_m$, $T_s$\} and 100~K in detail. We find that our $R(T)$ data can be described by $R= R_{\rm 0}+CT^\alpha$. Special care has been taken to eliminate the potential error introduced by thermal lag (see Supporting Information~S4). Figure~\ref{fig5} displays the temperature exponent $\alpha$ 
against pressure for all samples investigated in this study. Interestingly, the pressure-evolution of $\alpha$ can be separated into two distinct groups. In the thicker samples, $\alpha$ is suppressed from around 1.5 to around 0.8 as pressure increases (filled symbols in Fig.~\ref{fig5}). On the other hand, in the thinner samples, $\alpha$ is suppressed first and then it is gradually enhanced towards $\alpha=2$ at $p>$ 30~kbar (open symbols in Fig.~\ref{fig5}). Therefore, quasilinear $R(T)$ with $\alpha\approx1$ is only observed in the group of thicker samples. In other IBSs featuring a magnetic quantum critical point, $\alpha$ changes from unity at the quantum critical point to a higher value as the system is tuned away from the quantum critical point on the quantum disordered side~\cite{Kasahara2010, Shibauchi2014, Analytis2014}. Inspired by these studies, our results can be naturally interpreted as a demonstration of stronger magnetic fluctuations in thicker samples than in the thinner group where magnetism has also been substantially weakened.

Nematic fluctuations, which are expected to be significantly enhanced near the endpoint of the nematic phase, may also promote superconductivity, as suggested by both experimental and theoretical investigations~\cite{Mukasa2021, Kontani2010, Lederer2015, Lederer2017}. Similar to the result in FeSe$_{1-x}$Te$_x$ \cite{Mukasa2021}, in all samples studied here, $T_c$ is enhanced to a local maximum outside the nematic phase, not far from the extrapolated endpoint. However, since all six samples studied here exhibit very similar nematic phase boundaries, it is reasonable to assume that the character of nematic fluctuations does not vary significantly as the thickness reduces, in stark contrast to the greatly suppressed superconducting dome. Thus, we infer that while nematic quantum fluctuations can also promote superconductivity, such a promotion is limited in FeSe in the absence of a strong SDW phase. 

In summary, using thickness reduction and high pressure as tuning knobs, we systematically investigate the evolution of superconductivity, nematicity, and magnetism in thin FeSe with thicknesses ranging from 5~$\mu$m to 140~nm. We find that, as the thickness is reduced, the nematic phase boundary remains robust on $T$-$p$ phase diagrams, while the magnetically ordered phase shrinks significantly. Meanwhile, in all samples measured, a local maximum of $T_c$ is observed outside the nematic phase region, not far from the extrapolated nematic endpoint, suggesting that the nematic quantum fluctuations can enhance superconductivity. However, associated with the weakening of magnetism, the superconductivity is also weakened. The pressure-induced high-$T_c$ superconductivity no longer exists in our thinner samples, where magnetism is fragile. Our observations strongly suggest that nematic quantum fluctuations alone can only have a limited promotion of $T_c$, while magnetic fluctuations are pivotal for the stabilization of high-$T_c$ superconductivity.\\ 

\begin{center}{\bf METHODS}\end{center}
\textit{Crystal growths.}
High-quality single crystals of FeSe were grown by chemical vapor transport with a eutectic KCl-AlCl$_3$ mixture as the transport agent~\cite{Bohmer2013, Bohmer2016}. High-quality h-BN single crystal with the length of about 7~mm was grown under high pressure using a flux method. For detailed sample characterizations, see Supporting Information.\\ 
\\
\textit{Device integrated diamond anvil cell.}
 The diamond anvil cells used were equipped with two low side angle diamond anvils (Type IIas) with a culet diameter of 800~$\mu$m and a bevel of 1000~$\mu$m. Stainless steel gaskets with a 10~$\mu$m thick insulation layer were prepared following the method in Ref.~\cite{Thomasson1997}. The anvils inlaid in the piston were then patterned with micro electrodes by photo-lithography and PVD coating. Thin flakes of FeSe were exfoliated from single crystals using the ``blue tape" (from Nitto Denko Co.) and then attached to the silicone elastomer poly-dimethylglyoxime (PDMS, Gelfilm from Gelpak) stamp. Well-chosen flakes were then transferred onto the patterned electrodes. For 140~nm, 280~nm, 510~nm and 2.3~$\mu$m thick samples, a thin layer of h-BN was transferred onto the sample for encapsulation. The dimensions of the resulting devices were determined in a dual beam focused ion beam system (Scios 2 DualBeam from Thermo Scientific). High-purity glycerine was used as the pressure transmitting medium. The pressure was determined by the spectral measurement of ruby fluorescence at room temperature~\cite{Mao1986}.\\
\\
\textit{Measurements.}
Magneto-electrical resistance measurements were carried out in a Physical Property Measurement System by Quantum Design with a base temperature of 1.8~K and fields up to 14~T.\\

\begin{center}{\bf ASSOCIATED CONTENT}\end{center}
The Supporting Information is available.\\

Figures S1 -- S17 on the characterizations of FeSe and h-BN, ruby fluorescence spectra and additional transport data.

\begin{center}{\bf AUTHOR INFORMATION}\end{center}
{\bf Corresponding Authors}\\
$^*$Kwing To Lai. E-mail: ktlai@phy.cuhk.edu.hk\\
$^*$Swee K. Goh. E-mail: skgoh@cuhk.edu.hk\\

\noindent{\bf Author Contributions}\\
$^\ddagger$J.X. and X.L. contributed equally to this work.\\

\noindent{\bf Notes}\\
The authors declare no competing financial interest.

\begin{center}{\bf ACKNOWLEDGMENTS}\end{center}
The authors acknowledge insightful comments from Takasada Shibauchi, Anna B\"{o}hmer, Yajian Hu, Qun Niu, Fangdong Tang and Liyuan Zhang. The authors also thank Siu Kong Li, Chun Hei Lai and Man Hau Yeung for experimental supports and Sen Yang, Kangwei Xia for fruitful discussions on aspects of microfabrications. This work was supported by Research Grants Council of Hong Kong (CUHK 14301020, A-CUHK402/19), CUHK Direct Grant (4053461, 4053408, 4053410, 4053463), and the Key Research Platforms and Research Projects of Universities in Guangdong Province (Grant No. 2018KZDXM062). \\






\providecommand{\noopsort}[1]{}\providecommand{\singleletter}[1]{#1}%

\end{document}